\documentclass[aps,prb,twocolumn,superscriptaddress]{revtex4-2}
\usepackage{graphicx}
\usepackage{latexsym}
\usepackage{amssymb}
\usepackage{amsmath}
\usepackage{amsfonts}
\usepackage[vcentermath]{youngtab}
\usepackage{upgreek}
\usepackage{bm,dsfont}
\usepackage{multirow}
\usepackage{enumitem}
\usepackage{color}
\usepackage{hyperref}
\hypersetup{
colorlinks = true,
linkcolor = [rgb]{0.70,0.13,0.13},
citecolor = [rgb]{0.13,0.55,0.13},
urlcolor = [rgb]{0.25, 0.41, 0.88}}
\usepackage{makecell}

\newcommand{\dd}{\mathrm{d}}

\newcommand{\e}{\mathrm{e}}

\newcommand{\dsR}{\mathbb{R}}

\newcommand{\dsZ}{\mathbb{Z}}

\newcommand{\scE}{\mathcal{E}}

\newcommand{\scG}{\mathcal{G}}

\newcommand{\scV}{\mathcal{V}}

\newcommand{\Tr}{\operatorname{Tr}}

\newcommand{\mat}[1]{\left[\begin{matrix}#1\end{matrix}\right]}
\newcommand{\smat}[1]{\left[\begin{smallmatrix}#1\end{smallmatrix}\right]}
\newcommand{\eq}[1]{\begin{equation}#1\end{equation}}
\newcommand{\eqs}[1]{\begin{equation}\begin{split}#1\end{split}\end{equation}}
\newcommand{\eqnref}[1]{Eq.\,\eqref{#1}}
\newcommand{\figref}[1]{Fig.\,\ref{#1}}
\newcommand{\tabref}[1]{Tab.\,\ref{#1}}
\newcommand{\secref}[1]{Sec.\,\ref{#1}}

\newcommand{\tch}{\mathrm{tch}}
\newcommand{\std}{\mathrm{std}}

\begin{document}

\title{Machine Learning Renormalization Group for Statistical Physics}

\author{Wanda Hou}
\author{Yi-Zhuang You}
\affiliation{{Department of Physics, University of California at San Diego, La Jolla, CA 92093, USA}}
\date{\today}

\begin{abstract}
We develop a Machine-Learning Renormalization Group (MLRG) algorithm to explore and analyze many-body lattice models in statistical physics. Using the representation learning capability of generative modeling, MLRG automatically learns the optimal renormalization group (RG) transformations from self-generated spin configurations and formulates RG equations without human supervision. The algorithm does not focus on simulating any particular lattice model but broadly explores all possible models compatible with the internal and lattice symmetries given the on-site symmetry representation. It can uncover the RG monotone that governs the RG flow, assuming a strong form of the $c$-theorem. This enables several downstream tasks, including unsupervised classification of phases, automatic location of phase transitions or critical points, controlled estimation of critical exponents and operator scaling dimensions. We demonstrate the MLRG method in two-dimensional lattice models with Ising symmetry and show that the algorithm correctly identifies and characterizes the Ising criticality. 
\end{abstract}

\maketitle

\section{Introduction}

Renormalization group (RG) is an elegant conceptual framework and powerful computational method in statistical physics and quantum field theory. RG extracts the relevant features at every given scale by progressively coarsening the degrees of freedom in a physics system in hierarchies. Conventionally, real space RG relies on human physicists to design coarse-graining transformations based on their intuition of the physical system. In this research, we aim to explore the potential for artificial intelligence (AI) to learn optimal RG transformations automatically from energy models. Unsupervised machine learning is particularly well-suited for this task due to its ability to learn low-dimensional representations or relevant features from data and to remove noise and irrelevant features without supervision. This approach aligns with the goal of learning RG transformations, which aim to extract relevant features of a physical system by transforming fine-grained configurations to coarse-grained ones while preserving essential information and correlations.

Prior research has demonstrated that neural networks can learn to perform hierarchical feature extraction at the configuration level \cite{Koch-Janusz2018M1704.06279,Li2018N1802.02840,Efthymiou2018S1810.02372,Chung2019O1912.09005,Lenggenhager2018O1809.09632,Hu2020M1903.00804,Chung2021N2010.05703,Ron2021M2011.05567,Giataganas2022N2102.05219,Hu2022R2010.00029,Sheshmani2023C2203.07975}. However, a more fascinating aspect of RG is its ability to quantitatively analyze the flow of physics theory in the parameter space at the model level \cite{Di-Sante2022D2202.13268,2023PhRvB.108b4413U}. Therefore, this research aims to develop a novel machine-learning renormalization group (MLRG) method that can automatically formulate RG flow equations, discover RG monotones, propose effective theories, identify critical points, and estimate critical exponents, all starting from the symmetry and dimension of the physical system.

\begin{figure}[htbp]
\begin{center}
\includegraphics[scale=0.63]{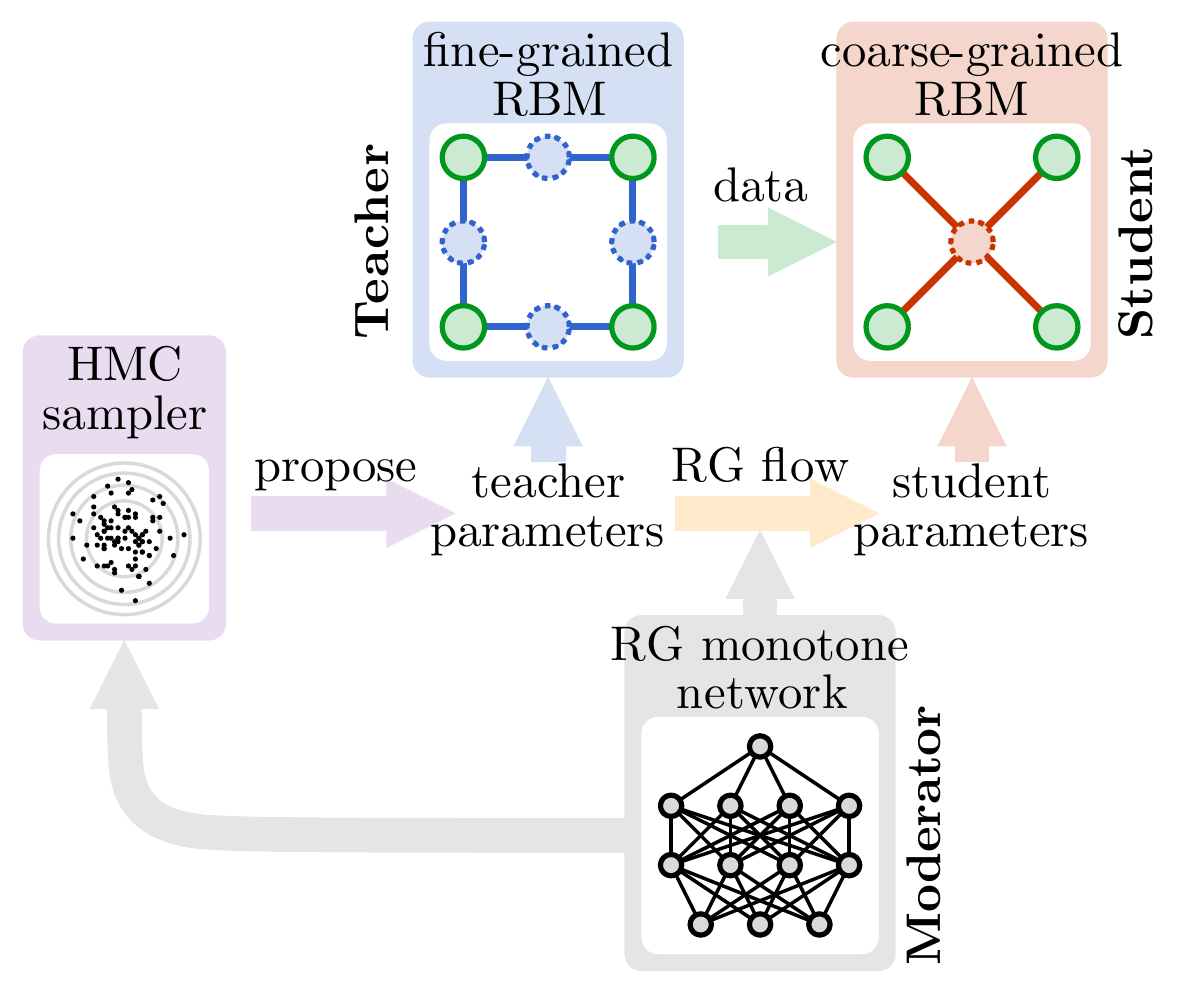}
\caption{High-level architecture of the MLRG algorithm. Three AI models are involved. The \emph{teacher} and \emph{student} models are both generative models based on RBM with different predefined neural network connects, representing the fine-grained and coarse-grained local energy models. The \emph{moderator} model is a discriminative model to predict the RG monotone. The moderator first guides a Hamiltonian Monte Carlo (HMC) sampler to propose the teacher model parameters that is most worth training, and then predicts the student model parameters following the RG flow. After both RBMs get their model parameters, the teacher generates the data to train the student. The gradient back-propagates from the student to the moderator to train the RG monotone network. After training, moderator gains good knowledge about the RG flow throughout the parameter space, which can then be used for various downstream tasks.}
\label{fig: architecture}
\end{center}
\end{figure}

In this study, we will focus on statistical mechanics models defined on regular lattices and develop machine learning algorithms to analyze them within the real space RG framework. Our proposed MLRG algorithm is depicted in \figref{fig: architecture}. We introduce two restricted Boltzmann machines (RBMs) \cite{Hinton1983O,Hinton2002T,Welling2004E,Hinton2012A} to model the local energy model at different scales on the fine-grained and coarse-grained lattices respectively. The fine-grained model serves as the teacher by generating samples of visible configurations, which are then used to train the coarse-grained model. The coarse-grained model acts as the student and learns its energy model to describe the training data provided by the teacher. We also introduce a third model, the RG monotone network, as a moderator that observes the teacher-student learning process and learns to predict how the model parameters of the coarse-grained model are related to those of the fine-grained model. This allows the moderator model to learn the RG flow and use its knowledge to guide a Hamiltonian Monte Carlo (HMC)  \cite{Duane1987H,Neal2011M,Betancourt2017A1701.02434} sampler to propose new model parameters that are most worth training. After training, we can use the machine-learned RG flow to identify RG fixed points in the parameter space and automate the analysis of physical properties at these fixed points.

The paper will be organized as follows. We will first introduce the MLRG algorithm in \secref{sec: method}, which includes (i) the teacher-student learning system \secref{sec: RSRG} to model the RG flow and (ii) the moderator \secref{sec: monotone} and HMC sampling \secref{sec: HMC} system to extract RG monotone and use it to guide the training. The teacher and student are modeled by equivariant RBMs, as formulated in \secref{sec: statmech}, whose point group representation choices are elaborated in \secref{sec: PG rep}. We then demonstrate the application of MLRG on 2D Ising models in \secref{sec: result}. \secref{sec: J assign} describes the problem setup. \secref{sec: RG flow} shows the machine-learned RG monotone and RG flow diagram. A few quantitative results are then presented, including the critical point \secref{sec: Jc}, the ground state degeneracy \secref{sec: GSD}, and the scaling dimension \secref{sec: Delta}. \secref{sec: Newton} explains how to use Newton's method to localize the RG fixed point. We summarize the advantage and limitations of MLRG and comment on its connections to related works in \secref{sec: summary}.

\section{Methodology}\label{sec: method}

\subsection{Statistical Mechanics Models}\label{sec: statmech}

We begin with a general definition of a statistical mechanics system on a lattice. Let the lattice be described by a graph $\scG=(\scV,\scE)$, where $\scV$ denotes the set of vertices (sites) and $\scE$ denotes the set of edges (bonds). On each site $i\in\scV$, we introduce a vector $s_i\in\dsR^n$ to characterize the on-site degrees of freedom, which are generally referred to as a \emph{spin}. The entire spin configuration $s$ on the lattice can be viewed as a map $s:\scV\to\dsR^n$. A central theme of equilibrium statistical physics is to model the probability distribution $p(s)\propto \e^{-E(s)}$ using a local energy function $E(s)$, such as
\eq{\label{eq: E general}
E_J(s)=\sum_{\langle ij\rangle\in\scE} \epsilon_J(s_i,s_j),}
where $\epsilon_J$ is a scalar function describing the energy associated with each pair of spins $(s_i, s_j)$ across the edge $\langle i j\rangle$. The subscript $J$ represents the collection of parameters that parametrize the energy model.

Symmetry plays a significant role in defining the spin degrees of freedom and constraining the energy function. Let $G$ be the internal symmetry group and $R$ be the point group of the lattice. Assuming $G$ and $R$ commute and form a direct product group $G\times R$, the on-site spin $s_i$ should form an $n$-dimensional linear representation of the $G\times R$ symmetry. Under the symmetry action $g\in G$ and $r\in R$, the spin $s_i$ transforms as
\eq{s_i\to (T_r\otimes T_g) s_{r(i)},}
where $T_g$ and $T_r$ are matrix representations of the internal $g$ and the point group $r$ symmetry transformations. $r(i)$ denotes the resulting site obtained from the original site $i$ by applying the point group transformation $r$.

The energy function $E_J(s)$ must be invariant under the internal and the lattice symmetry (including the point group and translation symmetries). This requires $\epsilon_J$ to satisfy the following symmetry constraint:
\eq{\label{eq: eps_J symm}
\epsilon_J(s_i,s_j)=\epsilon_J\big((T_r\otimes T_g) s_{r(i)}, (T_r\otimes T_g) s_{r(j)}\big)}
for all $(g,r)\in G\times R$. This means it is sufficient to define the energy function $\epsilon_J$ on a representative bond and use the lattice symmetry to carry it to every other bond on the lattice (assuming all bonds on the lattice are related by lattice symmetry transformations). 

\begin{figure}[htbp]
\begin{center}
\includegraphics[scale=0.65]{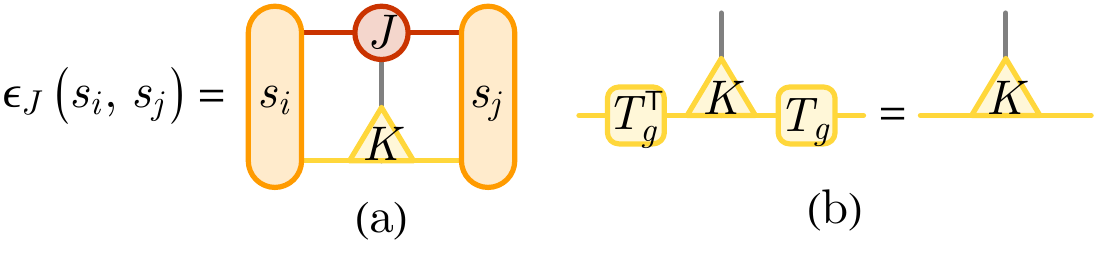}
\caption{(a) Tensor network representation of the bond energy function $\epsilon_J(s_i,s_j)$ on the representative bond. (b) The kernel $K$ is invariant under the internal symmetry $g\in G$.}
\label{fig: tensor}
\end{center}
\end{figure}

More explicitly, suppose the spin components $s_i^{a\alpha}$ are labeled by two indices $a$ and $\alpha$, separately indexing the representation space basis of the point group and internal symmetries. One possible design of an equivariant bond energy function $\epsilon_J$ on the representative bond is to use a tensor network:
\eq{\label{eq: bond energy TN}
\epsilon_J(s_i,s_j)=-J_{ab\gamma}K_{\alpha\beta}^{\gamma}s_i^{a\alpha}s_j^{b\beta},}
where repeated indices are automatically summed over, as illustrated in \figref{fig: tensor}(a). $K$ is a $G$-invariant tensor satisfying the symmetry constraint $T_{g}^{\alpha\alpha'}K_{\alpha'\beta'}^{\gamma}T_{g}^{\beta\beta'}=K_{\alpha\beta}^{\gamma}$ ($\forall g\in G$), as depicted in \figref{fig: tensor}(b). The tensor $J$ contains all parameters that determine the energy function. They can be viewed as coupling constants among different symmetry representations. Although \eqnref{eq: bond energy TN} does not yet represent the most general equivariant energy model, it is already sufficiently expressive if the representation space dimension $n$ is large enough. So we will not dive into more complicated equivariant neural network models \cite{Cohen2016G1602.07576,Kondor2018O1802.03690,Weiler201831807.02547,Cohen2018A1811.02017,Finzi2021A2104.09459,Lim2022W2205.07362} but adopt this tensor network design to construct the equivariant restricted Boltzmann machines later. 

In summary, given the internal symmetry group $G$ and the lattice graph $\scG$ (with the lattice symmetry given by the automorphism group of $\scG$), one can specify the on-site spin $s_i$ by its representation under the internal and point group symmetries. Statistical mechanical models are generally defined by symmetric energy functions $E_J(s)$, parametrized by $J$, as in \eqnref{eq: E general}. The RG analysis aims to determine how the model parameters $J$ effectively change across different scales.

\subsection{Real Space Renormalization Group}\label{sec: RSRG}

\begin{figure}[htbp]
\begin{center}
\includegraphics[scale=0.65]{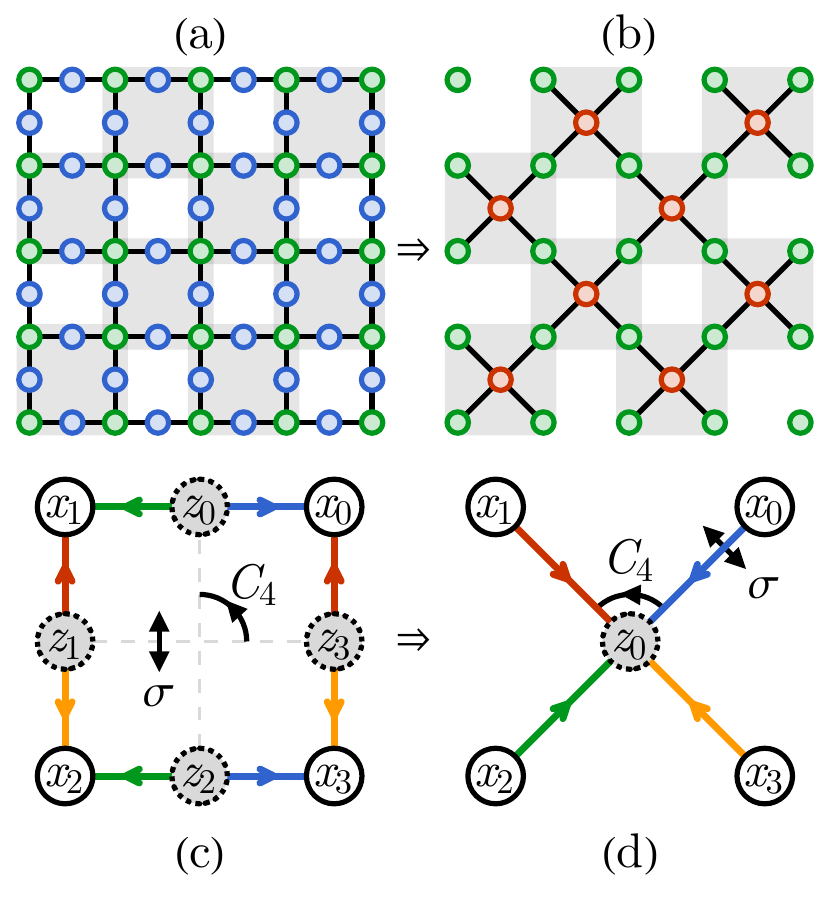}
\caption{(a) Fine-grained and (b) coarse-grained Lieb lattices. Shaded squares mark out the local blocks. In each block, the teacher and student RBMs are respectively defined on the  (c) square and (d) cross graphs. The arrows mark the bond directions. All bonds are related by $C_{4v}$ point group transformations, which include the four-fold rotation $C_4$ and the mirror reflection $\sigma$.}
\label{fig: lattice}
\end{center}
\end{figure}

For concreteness, we will focus on two-dimensional lattice models. In particular, we will consider the Lieb lattice \cite{Lieb1989T} (bond-intercalated square lattice) as shown in  \figref{fig: lattice}(a,b). The Ising model defined on the Lieb lattice is physically equivalent to the simple square lattice but the Lieb lattice is more convenient for describing real-space RG schemes. Extending our approach to other lattices and higher dimensions is possible in general.

Starting with an energy model defined on the Lieb lattice, our RG scheme can be described as follows: (i) Divide the lattice into overlapping $2\times 2$ blocks as in \figref{fig: lattice}(a). (ii) Within each block, replace the local energy model on a square graph \figref{fig: lattice}(c) with a new model on a cross graph \figref{fig: lattice}(d), such that their marginal distributions on the corner spins match as closely as possible. (iii) Embed the new local energy model back into the original lattice. The new lattice \figref{fig: lattice}(b) becomes a coarse-grained Lieb lattice, with lattice constant enlarged by $\sqrt{2}$. Repeating the procedure recursively, the energy model parameters will be renormalized to a larger and larger lattice scale.

The key objective of this RG scheme is to learn the new local energy model. For this purpose, we view the square- and cross-graph local energy models as two restricted Boltzmann machines (RBMs) \cite{Hinton1983O,Hinton2002T,Welling2004E,Hinton2012A}. We will call the RBM on the fine-grained square graph the \emph{teacher} machine, in which the corner spins $x_i$ are the visible variables and the decorated spins $z_i$ are the hidden variables, see \figref{fig: lattice}(c). Its energy model reads:
\eq{\label{eq: E tch}
E_{J_\tch}(x,z)=\sum_{\langle ij\rangle\in\scE_\text{square}}\epsilon_{J_\tch}(z_i,x_j).
}
The RBM on the coarse-grained cross graph will be called the \emph{student} machine, in which the corner spins $x_i$ are still the visible variables, but there is only one hidden spin $z_0$ at the center, see \figref{fig: lattice}(d). Its energy model is :
\eq{\label{eq: E std}
E_{J_\std}(x,z)=\sum_{\langle ij\rangle\in\scE_\text{cross}}\epsilon_{J_\std}(x_i,z_j).
}
The teacher and student RBMs are separately parametrized by $J_\tch$ and $J_\std$. In both \eqnref{eq: E tch} and \eqnref{eq: E std}, the bond energy function $\epsilon_J(x_i,x_j)$ is defined on a representative bond along the $i\to j$ direction using the tensor network model \eqnref{eq: bond energy TN}. In \figref{fig: lattice}(c,d), the representative bonds are colored in blue, and arrows indicate the bond directions. 

Both the square and cross graphs respect the $C_{4v}$ point group symmetry, which is generated by a four-fold rotation $C_4$ and a mirror reflection $\sigma$, see \figref{fig: lattice}(c,d). The mirror reflection $\sigma$ is always assigned to preserve the representative bond (in blue). It only imposes symmetry constraints on the bond energy model $\epsilon_J$ as required by \eqnref{eq: eps_J symm}. The four-fold rotation $C_4$ further takes the representative bond to other bonds (of other colors), under which the bond energy model will change equivariantly following \eqnref{eq: eps_J symm}. This way, the RBMs will respect the internal $G$ and point group $R=C_{4v}$ symmetries by design. 

The objective is to train the student RBM to learn the coarse-grained local energy model from the visible spin configurations generated by the teacher RBM. Given the teacher model parameters $J_\tch$, we optimize student model parameters $J_\std$ by minimizing the Kullback-Leibler (KL) divergence between the visible variable distributions of the two models,
\eq{\label{eq: DKL def}
D_\text{KL}(p_\tch(x)\Vert p_\std(x))=\sum_{x}p_\tch(x)\log\frac{p_\tch(x)}{p_\std(x)},}
where the marginal distributions $p_\tch(x)$ and $p_\std(x)$ are defined by tracing out the hidden spins:
\eqs{p_\tch(x)&=\frac{\sum_{z}\e^{-E_{J_\tch}(x,z)}}{\sum_{x',z}\e^{-E_{J_\tch}(x',z)}},\\
p_\std(x)&=\frac{\sum_{z}\e^{-E_{J_\std}(x,z)}}{\sum_{x',z}\e^{-E_{J_\std}(x',z)}}.}
Although directly evaluating the KL divergence in \eqnref{eq: DKL def} is not tractable, the optimization can be performed by stochastic gradient descent following the standard contrastive divergence (CD) \cite{Hinton2002T} training technique for the RBM. 

In this way, the student machine automatically learns the coarse-grained model parameters $J_\std$ given the fine-grained model parameters $J_\tch$ of the teacher machine. This contrasts with the conventional real space RG approach, where human physicists must design the coarse-graining rule that maps the visible spin configuration $x$ to a hidden spin $z_0(x)$ in each block. Our approach differs in two aspects: (i)  The coarse-grained variable $z_0$ is no longer a deterministic map of $x$ but is defined in a probabilistic manner via a conditional distribution $p_\std(z_0|x)$ given by the student machine. (ii) The conditional distribution $p_\std(z_0|x)$ is not specified by humans but is learned from the data. This allows the student machine to develop the optimal RG transformation within its available representation space dimensions, extracting the most relevant features $z_0$ from the visible configurations $x$ without supervision.

After training, we can replace the teacher model parameters $J_\tch$ with the student model parameters $J_\std$ and move on to train the next-generation student. Therefore, generation by generation, the teacher-student learning iteration will trace the RG flow of $J$ in the parameter space.

\subsection{Choosing Point Group Representations}\label{sec: PG rep}

To demonstrate how well the student RBM can approximate the teacher RBM, we consider a concrete example based on the two-dimensional Ising model on the square lattice. In this case, the internal symmetry is $G=\dsZ_2$, and the point group symmetry is $R=C_{4v}$. Then every spin variable (either visible or hidden) in both RBMs will be specified as a representation of $\dsZ_2\times C_{4v}$. The  $\dsZ_2$ group only has one non-trivial representation, i.e., the odd (signed) representation that transforms as $s\to-s$. So we will assume every component of the spin to transform as this odd representation under the $\dsZ_2$ Ising symmetry. The $C_{4v}$ point group has richer irreducible representations, summarized in \tabref{tab: D4h}. The expressive power of the RBM depends on the choice of the $C_{4v}$ point group representations for each spin, which will be elaborated in the following.

\begin{table}[htbp]
\centering
\begin{tabular}{c|cccc}
irrep & dim & transforms as &  $T_{C_4}$ & $T_{\sigma}$ \\
\hline
$\mathsf{A}_1$ & $1$ & $1$ & $\smat{1}$ & $\smat{1}$ \\
$\mathsf{A}_2$ & $1$ & $xy(x^2-y^2)$ & $\smat{1}$ & $\smat{-1}$ \\
$\mathsf{B}_1$ & $1$ & $x^2-y^2$ & $\smat{-1}$ & $\smat{1}$ \\
$\mathsf{B}_2$ & $1$ & $xy$ & $\smat{-1}$ & $\smat{-1}$ \\
$\mathsf{E}$   & $2$ & $(x,y)$ & $\smat{0&-1\\1&0}$ & $\smat{1&0\\0&-1}$ \\
\end{tabular}
\caption{Irreducible representations of the $C_{4v}$ group. The columns shows their dimensions, Cartesian products ($x$-axis along the representative bond), and transformation matrices $T_r$ for $r=C_4,\sigma$ (the two generators of $C_{4v}$).}
\label{tab: D4h}
\end{table}

A single Ising spin in the conventional Ising model corresponds to a $\dsZ_2$-odd variable carrying $\mathsf{A}_1$ representation, which does not transform under the point group symmetry at all. However, under RG, the coarse-grained spin (such as the $z_0$ spin in the student RBM) can carry an enlarged point group representation. From \figref{fig: lattice}(c,d), it is clear that the RG procedure is effectively merging the four hidden spins in the teacher model (square graph) to one hidden spin in the student model (cross graph). Therefore the hidden spin $z_0$ in the student model must contain more internal structures and carry orbital angular momentum (i.e., non-trivial representation of $C_{4v}$) to resolve the complication of inhomogeneous hidden spin configurations in the teacher model.

For example, by saying that a $\dsZ_2$ spin variable $z_0$ carries the $\mathsf{A}_1\oplus \mathsf{E}$ representation, we imply that $z_0\in\dsR^3$ is a three-component vector, consisting of three Ising variables: one Ising variable forms the $\mathsf{A}_1$ representation and remains invariant under lattice rotations, and two Ising variables form the two-dimensional $\mathsf{E}$ representation that can rotate as a vector. Then the $C_4$ and $\sigma$ transformations are represented as
\eq{T_{C_4}z_0=\smat{1&0&0\\0&0&-1\\0&1&0}z_0, \quad T_{\sigma}z_0=\smat{1&0&0\\0&1&0\\0&0&-1}z_0,}
according to the transformation matrices listed in \tabref{tab: D4h}.

Suppose we start from a fine-grained teacher model with both the visible and hidden spins as ordinary Ising spins in the $\mathsf{A}_1$ representation, i.e., $x_i,z_i\in\{\pm 1\}$. The parameter tensor $J_\tch$ will contain only one component given the symmetry constraints. According to the square graph in \figref{fig: lattice}(c), the teacher model is described by the energy function below
\eq{E_{J_\tch}(x,z)=-J_\tch\sum_{i=0}^{3}z_i(x_i+x_{(i+1)\mathrm{mod} 4}).}
The hidden spins $z_i$ can be immediately traced out (marginalized) in the partition function, leaving us an effective model for visible spins $x_i$, described by
\eq{\label{eq: Etch toy}
E_{J_\tch}(x)=-J_\text{eff}(x_0x_1+x_1x_2+x_2x_3+x_3x_0),}
where $J_\text{eff}=\frac{1}{2}\log\cosh(2J_\tch)$ is an effective coupling that depends on $J_\tch$. Correspondingly, the visibile spin discributions is $p_\tch(x)=Z^{-1}\e^{-E_{J_\tch}(x)}$.

To build a coarse-grained student model to approximate $p_\tch(x)$, the first step is to specify the choice of the point group representation for the coarse-grained variable $z_0$ in the student RBM. For example, if $z_0$ is chosen to carry the $\mathsf{A}_1\oplus \mathsf{E}$ representation with totally three components as $z_0=(z_0^1,z_0^2,z_0^3)$, where each component is an $\dsZ_2$ Ising variable that $z_0^a\in\{\pm 1\}$, the student model can be written as (see \figref{fig: lattice}(d) for the connectivity of these variables)
\eqs{\label{eq: Estd toy}
E_{J_\std}(x,z)=&-J_\std^{\mathsf{A}_1\mathsf{A}_1}(x_0+x_1+x_2+x_3)z_0^{1}\\
&-J_\std^{\mathsf{A}_1\mathsf{E}}\big((x_0-x_2)z_0^{2}+(x_1-x_3)z_0^{3}\big).}
The fluctuation of the $\mathsf{A}_1$ representation $z_0^1$ mediates an equal amount of positive correlation $\langle x_ix_j\rangle$ between every pair of visible spins, which is indeed the dominant correlation pattern in a ferromagnetic Ising model. However, to better match the teacher model $p_\tch(x)$, the $\langle x_0x_2\rangle$ and $\langle x_1x_3\rangle$ correlations should be further weakened relative to others because there is no direct coupling between these diagonal spins in the original teacher model \eqnref{eq: Etch toy}. This is where the $\mathsf{E}$ representation plays a role, as the $z_0^2$ and $z_0^3$ fluctuations will mediate some {negative} correlation in $\langle x_0x_2\rangle$ and $\langle x_1x_3\rangle$ due to the minus signs in \eqnref{eq: Estd toy}, which helps to bring the student model's visible distribution $p_\std(x)$ closer to that of the teacher model $p_\tch(x)$. This argument explains why including more point group representations to the hidden spin in the student model is generally helpful to improve its expressive power.

\begin{figure}[htbp]
\begin{center}
\includegraphics[scale=0.65]{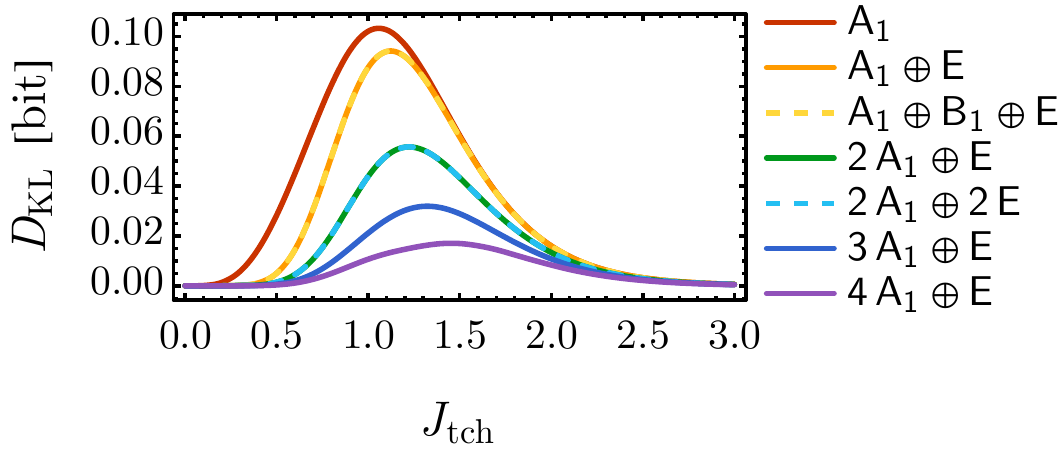}
\caption{Minimal KL divergence $D_\text{KL}(p_\tch(x)\Vert p_\std(x))$ (in the unit of bit) vs. the parameter $J_\tch$ of the teacher RBM, for different choices of point group representations of $z_0$ in the student RBM.}
\label{fig: DKL}
\end{center}
\end{figure}

Within this setup, we tested the performance of a series of student RBMs, built with different choices of the point group representation for the hidden spin $z_0$. The performance is evaluated by the minimal KL divergence $D_\text{KL}(p_\tch(x)\Vert p_\std(x))$ achieved after optimizing the student machine. In \figref{fig: DKL}, we show the minimum KL divergence over a large range of the parameter $J_\tch$ of the teacher model under different representation choices of $z_0$. The general  representations are constructed by combining irreducible representations $\mathsf{A}_1$, $\mathsf{B}_1$, and $\mathsf{E}$. The other two irreducible representations $\mathsf{A}_2$ and $\mathsf{B}_2$ are not used in this specific calculation because they will not couple to the $x_i$ spins in the $\mathsf{A}_1$ representation due to the constraint from the mirror reflection symmetry $\sigma$. But in more general cases, when the representation of $x_i$ is larger, all irreducible representations can appear on $z_0$ in principle. This calculation quantitatively demonstrates that the student RBM can learn to approximate the visible spin distribution from the teacher RBM. The approximation can be progressively improved by introducing larger representations on the hidden spin $z_0$, even though an exact match might only be achievable in the large representation dimension limit $n\to \infty$.

In practice, the RBM can only handle a finite representation space dimension $n$. So we have to truncate the representation space at a maximum dimension $n_\text{max}$. This will introduce inaccuracy in each RG step, but the hope is that the error introduced by the truncation will become irrelevant under RG flow, such that the truncation only affect the RG fixed point behavior in a controllable manner. This is also the underlying assumption of many real space RG schemes, where each RG step does not need to preserve the partition function exactly.

As we carry out the MLRG procedure by recursively training the student RBM by the teacher RBM and replacing the teacher with the trained student, the RG flow is supposed to bring us (close) to some fixed point RBM model, at which we can further investigate the universal properties (such as critical exponents and operator scaling dimensions) of the RG fixed point. We will provide numerical evidence to show that the MLRG algorithm can locate RG fixed points and estimate their universal properties more accurately when the representation dimension $n$ gets larger, such that it could offer a useful and controllable numerical method to automatically map out phase diagrams and study critical phenomena in statistical physics models without human supervision.

\subsection{Learning the RG Monotone}\label{sec: monotone}

While the above teacher-student learning approach for real space RG is appealing, it faces a serious challenge in actual training. This challenge comes from the stochastic nature of training the RBM, such that the student RBM parameter $J_\std$ will always fluctuate around its optimum in every RG step. This inevitably injects random noise into the entire RG flow, causing the model to perform random walks in the parameter space. Since the RG flow near the critical point (i.e., unstable RG fixed point) is particularly sensitive to small perturbations, stochastic RG flow will almost always miss the critical point. Therefore, without addressing the problem of parameter space random walk, the above naive MLRG algorithm is useless for studying any critical phenomena.

To address this challenge, we introduce a third AI system, called the \emph{moderator}, which operates outside the teacher-student system. The moderator’s objective is to monitor the stochastic RG flow over time in many different scenarios and learn the underlying deterministic RG flow throughout the entire parameter space. The key idea here is to assume the existence of an RG monotone $C(J)\in\dsR$, which is a real scalar function of the RBM model parameters $J$, such that the RG flow can be formulated as a gradient flow of the RG monotone $\dd J/\dd\ell=-\nabla_{J}C(J)$, where $\ell$ parametrize the RG step. This is the strongest form of the $c$-theorem \cite{Zomolodchikov1986, Barneshep-th/0408156, Friedan0910.3109, Komargodski1107.3987}. Instead of trying to construct such RG monotone by humans, we introduce a feed-forward neural network $C_\theta(J)$, called the RG monotone network, to model the function $C(J)$ and train it jointly with the RBMs in the RG flow. Here $\theta$ denotes the collection of all parameters in the neural network.

The training starts from a random choice of the teacher model parameters $J_\tch$ in the parameter space. The moderator takes the initial condition $J(\ell=0)=J_\tch$ and evolves $J(\ell)$ from $\ell=0$ to $\ell=1$ by solving the RG equation
\eq{\label{eq: RG flow}
\frac{\dd J}{\dd\ell}=-\nabla_JC_\theta(J),}
following the gradient signal provided by the RG monotone network $C_\theta$. The neural ordinary differential equation (neuralODE) technique \cite{Chen2018N1806.07366,Grathwohl2018F1810.01367} is employed here to enable gradient backpropagation through the ODE solver. The moderator then passes the solution to the student RBM as its model parameters $J_\std=J(\ell=1)$. The teacher RBM then starts sampling visible spin configurations and sends them to the student RBM. The student RBM receives the training data and evaluates the KL divergence in \eqnref{eq: DKL def} as the total loss function. All models are trained jointly by minimizing the KL divergence using the CD training technique. The loss function gradient will eventually back-propagate to $\theta$ to update the RG monotone network. The training is performed at many random choices of initial parameters $J_\tch$, such that a large amount of training data can be aggregated to optimize the RG monotone throughout the parameter space. 

Although the nature of the training is still stochastic, the random noise will be averaged out in fitting the RG monotone, such that after training, the optimal fit $C_{\theta_*}(J)$ can be used to generate a deterministic RG flow following \eqnref{eq: RG flow}. In this way, the random walk behavior in the parameter space can be avoided, which enables the MLRG algorithm to locate RG fixed points accurately.

\subsection{Sampling the Parameter Space}\label{sec: HMC}

Next, we will discuss how to sample $J_\tch$ more efficiently to speed up the training. Since the RBM model parameters $J$ live in high-dimensional parameter space, uniform sampling might not be an efficient strategy. We propose an importance sampling strategy that focuses on RG fixed points. 

The sampler makes use of the knowledge about the RG monotone to sample the parameters $J$ from the following probability distribution
\eq{\label{eq: p(J)}
p(J)\propto \e^{-\beta \Vert \nabla_J C_\theta(J)\Vert^2},}
with some inverse-temperature $\beta$ as a hyperparameter. We set $\beta=0$ initially, such that the sampler will propose model parameters $J$ uniformly in the parameter space. The RG monotone network gradually accumulates knowledge about the RG flow as the training progresses. RG fixed points emerge as local saddle points where the gradient $\nabla_J C_\theta(J)=0$ vanishes. We then gradually increase $\beta$, so the sampler will be biased to sample more around RG fixed points (including both stable and unstable). This design encourages the RG monotone network to be trained more intensively near RG fixed points, such that the fixed point location can be estimated more accurately, a desirable feature for the automatic discovery of critical points (phase transitions) in statistical mechanics systems. 

Since $J$ varies continuously, the Hamiltonian Monte Carlo (HMC) approach \cite{Duane1987H,Neal2011M,Betancourt2017A1701.02434} becomes a natural choice of the sampling method. HMC is a Markov chain Monte Carlo algorithm that uses Hamiltonian dynamics to efficiently propose moves in the parameter space of the target distribution $p(J)$. It is a powerful method for sampling complex distributions of continuous variables in high-dimensional space. We implement the method using multiple HMC samplers with replica exchange to mitigate the possibility of getting trapped at a single fixed point. 

In summary, the RG monotone network $C_\theta(J)$ plays a fundamental role in the MLRG algorithm, as illustrated in \figref{fig: architecture}. It first guides the HMC sampler to propose new teacher model parameters $J_\tch$ following the probability distribution in \eqnref{eq: p(J)} and then predicts the student model parameters $J_\std$ by solving the RG flow differential equation in \eqnref{eq: RG flow}. In this sense, it indeed serves as a moderator to moderate the teaching-learning process. In return, the RG monotone network gets trained when the student RBM learns to generate similar spin configurations as those generated by the teacher RBM. It is worth mentioning that the HMC sampled teacher model parameters $J_\tch$ are detached from the computational graph, such that the RG monotone network parameters $\theta$ will only receive gradient signals from the student model parameters $J_\std$. In this way, the training of the RG monotone will not be affected by the HMC sampling.

\section{Results and Analysis}\label{sec: result}

\subsection{Symmetry Assignment and Coupling Parameters}\label{sec: J assign}

To apply the MLRG method, we focus on two-dimensional lattice models with Ising symmetry. We do not need to specify any particular Ising model Hamiltonian, as the MLRG can automatically explore all possible models that are consistent with the internal and lattice symmetry, given the on-site symmetry representation.

In the following, we will always take the $G=\dsZ_2$ internal symmetry and the $R=C_{4v}$ point group symmetry. Regarding the symmetry representation of on-site spins, we consider that every spin (whether visible or hidden) is odd under $\dsZ_2$ and carries $\mathsf{A}_1$ or $\mathsf{A}_1\oplus\mathsf{E}$ or $2\mathsf{A}_1\oplus\mathsf{E}$ representation under $C_{4v}$. The choice of these point group representations is based on our experience that they are the most efficient representation within their respective representation dimensions in minimizing the KL divergence between teacher and student RBMs, as shown in \figref{fig: DKL}. The above symmetry assignment is summarized in \tabref{tab: symm Ising}, which is all we need to set up the equivariant RBMs and prepare the MLRG model for training.

\begin{table}[htbp]
    \centering
    \begin{tabular}{ccc}
         & symmetry & representation \\
    \hline
    $G$  & $\dsZ_2$ & $-1$ \\
    $R$  & $C_{4v}$ & $\mathsf{A}_1, \mathsf{A}_1\oplus\mathsf{E}, 2\mathsf{A}_1\oplus\mathsf{E}$
    \end{tabular}
    \caption{Settings of symmetries and spin representations for the two-dimensional Ising MLRG.}
    \label{tab: symm Ising}
\end{table}

For $\dsZ_2$ spins, the (representative) bond energy function in \eqnref{eq: bond energy TN} reduces to the following form \eq{
\epsilon_J(s_i,s_j)=-J_{ab}s_i^{a}s_j^{b},
}
where $a,b$ label the basis of the $C_{4v}$ symmetry representation. The internal symmetry representation labels $\alpha,\beta$ in \eqnref{eq: bond energy TN} are omitted because the $\dsZ_2$ group only has one non-trivial irreducible representation. The coupling $J$ is a matrix that can be parametrized as follows under different choices of the point group representations:
\begin{itemize}
\item $\mathsf{A}_1$ representation (1-dimension, 1 coupling parameter)
\eq{J=\mat{J_0},}
\item $\mathsf{A}_1\oplus\mathsf{E}$ representation (3-dimensional, 5 coupling parameters)
\eq{\label{eq: J A1E}
J=\left[\begin{array}{c|cc}
J_0&J_1&0\\
\hline
J_2&J_3&0\\
0&0&J_4
\end{array}\right],}
\item $2\mathsf{A}_1\oplus\mathsf{E}$ representation (4-dimensional, 10 coupling parameters)
\eq{J=\left[\begin{array}{c|c|cc}
J_0&J_1&J_2&0\\
\hline
J_3&J_4&J_5&0\\
\hline
J_6&J_7&J_8&0\\
0&0&0&J_9
\end{array}\right].}
\end{itemize}
In the above matrices, we use lines to separate different irreducible representations of the $C_{4v}$ symmetry. Some matrix elements are zero because of the restriction imposed by the mirror reflection symmetry $\sigma$, as required by \eqnref{eq: eps_J symm} in general. In all cases, the parameter $J_0$ always denotes the Ising coupling between the first $\mathsf{A}_1$ representations, directly connected to the bare Ising coupling in the lattice model.

\subsection{RG Monotone and RG Flow}\label{sec: RG flow}

\begin{figure}[htbp]
\begin{center}
\includegraphics[scale=0.65]{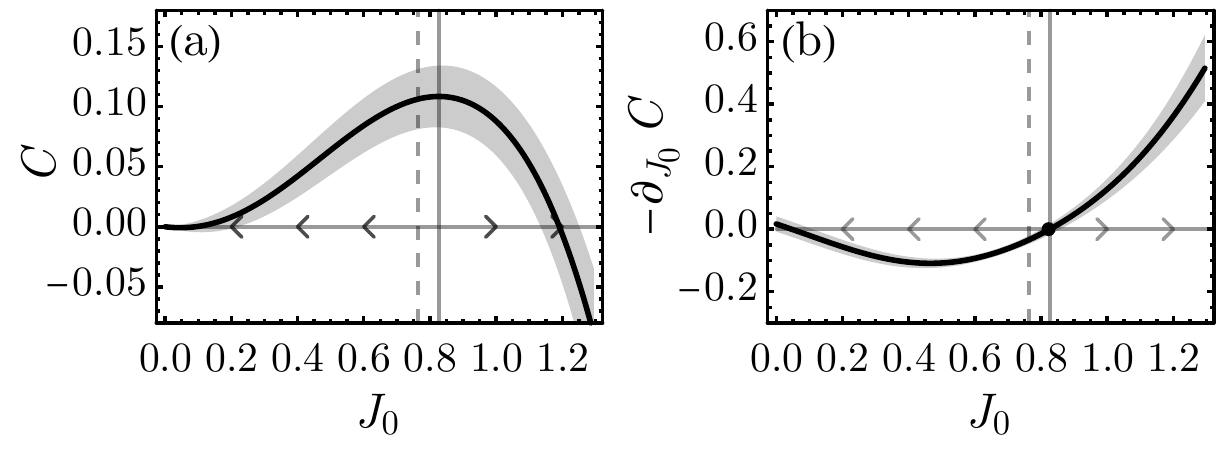}
\caption{(a) RG monotone $C(J_0)$ and (b) its negative gradient $-\partial_{J_0}C(J_0)$ (beta function), as functions of the unique coupling parameter $J_0$ in the $\mathsf{A}_1$ representation. The shaded area indicates the two-standard deviation uncertainty range, estimated based on ten independently trained MLRG models. The vertical solid line marks the estimated critical point $J_{0*}$, while the exact critical point $J_{0c}$ is displayed as the vertical dashed line. Arrows indicate the RG flow directions.}
\label{fig: A1mono}
\end{center}
\end{figure}

Choosing the smallest point group representation $\mathsf{A}_1$, the coupling matrix $J$ is parametrized by a single variable $J_0$. In this case, the RG flow is simply defined in the one-dimensional parameter space. The RG monotone $C(J_0)$, determined through the MLRG method, is depicted in \figref{fig: A1mono}(a). A local maximum of the RG monotone is observed at $J_{0*}=0.82\pm0.02$. Since the RG flow is designed to follow the gradient descent trajectory of the RG monotone as in \eqnref{eq: RG flow}, it implies that the parameter $J_0$ will depart from the unstable fixed point $J_{0*}$ and flow towards one of the two stable fixed points, either $J_0=0$ or $J_0\to \infty$. The RG flow can also be understood from the plot of $-\partial_{J_0}C$ in \figref{fig: A1mono}(b), which corresponds to the beta function in the context of RG theory.

The maximal position $J_{0*}:=\mathrm{argmax}_{J_0}C(J_0)$ of the RG monotone $C(J_0)$ provides an estimation of the Ising critical point that separates the paramagnetic phase ($J_0=0$) and the ferromagnetic phase ($J_0\to \infty$). The exact Ising critical point of the Lieb lattice model is located at
\eq{J_{0c}=\frac{1}{2}\mathrm{arccosh}(1+\sqrt{2})\approx 0.7643.}
The estimation $J_{0*}=0.82\pm 0.02$ is close to but still deviates from the exact value $J_{0c}$. This is because the dimension of the $\mathsf{A}_1$ representation is small, which limits the ability of the student RBM to approximate the teacher RBM in the MLRG algorithm. We should expect the estimation to be improved as the on-site spin involves larger representations of the point group.

\begin{figure}[htbp]
\begin{center}
\includegraphics[scale=0.65]{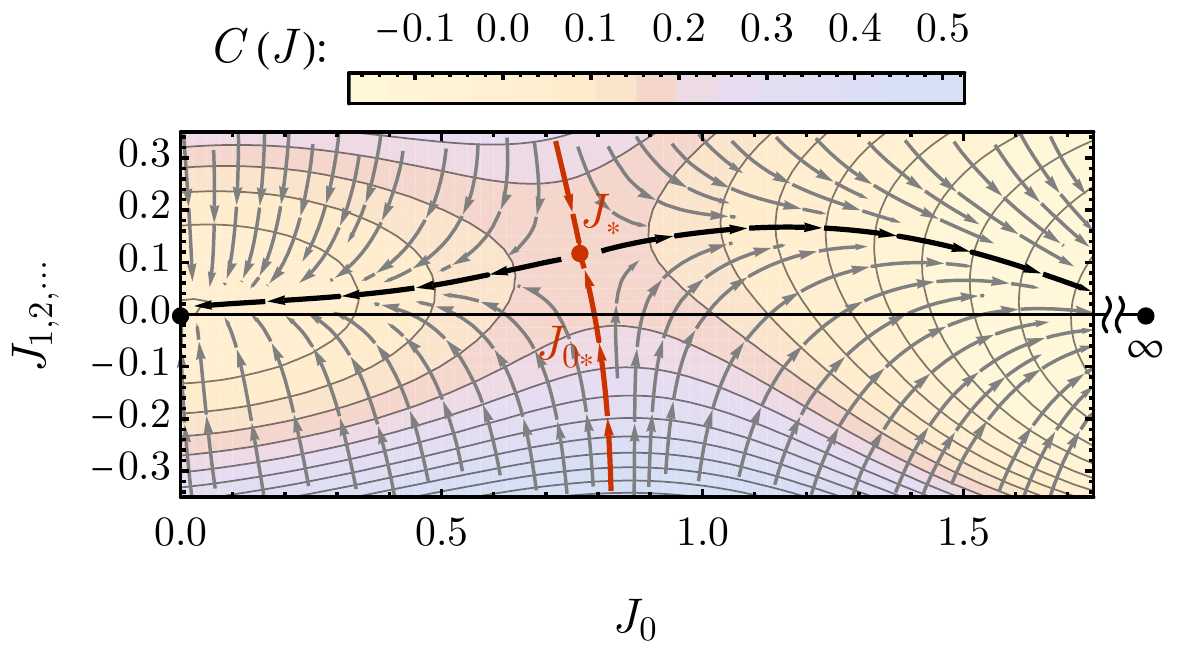}
\caption{RG flow diagram for models with $\mathsf{A}_1\oplus \mathsf{E}$ representation. $J_0$ parametrize the Ising coupling between $\mathsf{A}_1$ spins. $J_{1,2,\cdots}$ parametrize the remaining coupling in \eqnref{eq: J A1E} in their most relevant direction. The background color (and contours) indicate the RG monotone $C(J)$. The arrows trace out the RG flow directions. The stable (unstable) fixed points are marked by black (red) dots.}
\label{fig: RG flow}
\end{center}
\end{figure}

Going beyond the $\mathsf{A}_1$ representation, the RG flow will be defined in high-dimensional parameter space. To visualize the RG monotone $C(J)$, we separate the parameter $J_0$ from the remaining parameters $J_{1,2,\cdots}$ in the coupling matrix $J$. We always initialize the RG flow by setting $J_0$ to the bare Ising coupling of the lattice model and allowing $J_{1,2,\cdots}$ to be generated under the RG flow. On the sub-manifold spanned by $J_0$ and the most relevant flow direction of $J_{1,2,\cdots}$ (which denotes a particular linear combintation of $J_1,J_2,\cdots$ parameters along which the gradient of the RG monotone is maximal), we can plot the RG monotone $C(J)$ obtained by the MLRG method, as well as its gradient directions (RG flow directions). An example is shown in \figref{fig: RG flow}, which is obtained by training with the $\mathsf{A}_1\oplus \mathsf{E}$ on-site representation.

The MLRG algorithm learns the RG monotone $C(J)$ throughout the parameter space, based on which the RG flow diagram can be obtained. As we tune the bare coupling $J_0$ to the critical point $J_{0*}$ along the horizontal axis, the RG flow takes us to the true RG fixed point $J_{*}$ away from the horizontal axis. The parameter $J_{*}$ defines a statistical mechanics model approximating the Ising conformal field theory (CFT). The approximation is expected to be better with larger on-site point group representations. 

\subsection{Determining the Critical Point}\label{sec: Jc}

\begin{figure}[htbp]
\begin{center}
\includegraphics[scale=0.65]{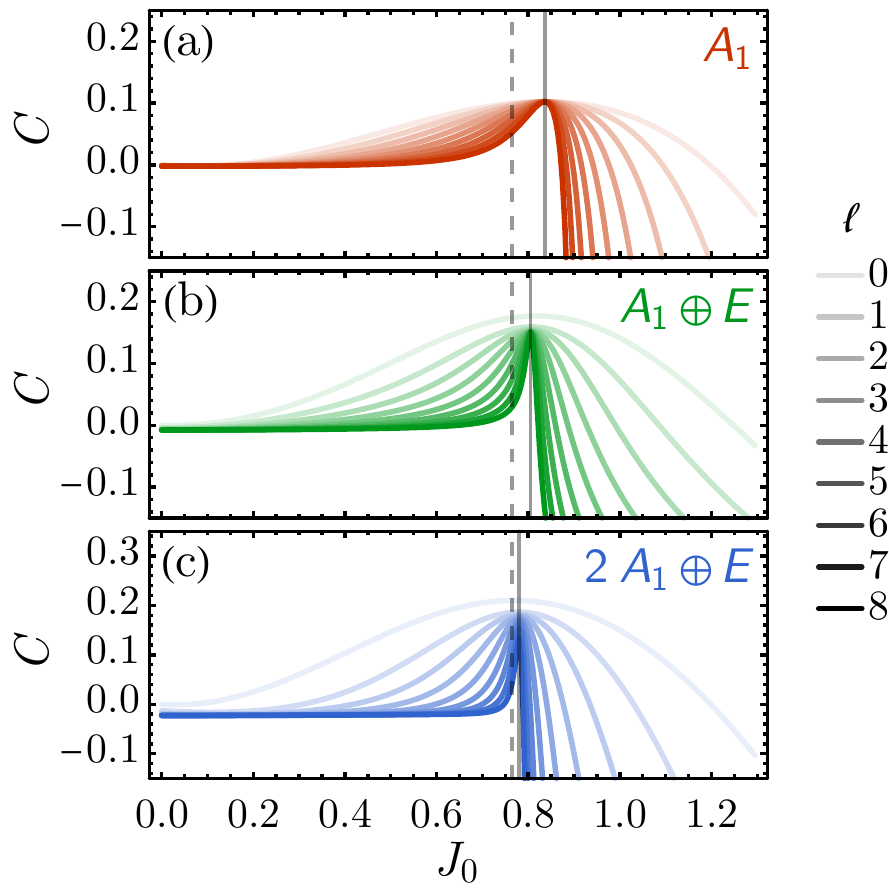}
\caption{Flow of the RG monotone $C(J(J_0,\ell))$ starting from given $J_0$ parameters, under the representation choice of (a) $\mathsf{A}_1$, (b) $\mathsf{A}_1\oplus \mathsf{E}$, and (c) $2\mathsf{A}_1\oplus\mathsf{E}$. $\ell$ labels the steps of RG flow. The vertical solid line marks the estimated critical point $J_{0*}$, while the exact critical point $J_{0c}$ is displayed as the vertical dashed line.}
\label{fig: monoflow}
\end{center}
\end{figure}

To estimate the Ising critical point $J_{0*}$, we start by initializing the coupling matrix $J$ with a given $J_0\in\dsR$ and $J_{1,2,\cdots}=0$, denoted as
\eq{\label{eq: J init}
J(J_0,\ell=0)=\mat{J_0&0&\cdots\\0&0&\cdots\\ \vdots&\vdots&\ddots}.}
Then we flow the coupling matrix $J$ by solving the RG equation $\dd J/\dd\ell=-\nabla_J C(J)$ from $\ell=0$ to some large $\ell$. We denote the solution as $J(J_0,\ell)$ as it depends on the initial condition $J_0$ and the RG scale $\ell$. Under our RG scheme, the linear system size will be effectively enlarged by $2^{\ell/2}$ at the RG scale $\ell$. \figref{fig: monoflow} shows the RG monotone $C(J(J_0,\ell))$ as a function of different initial value of $J_0$ at different RG scales $\ell$.

\begin{figure}[htbp]
\begin{center}
\includegraphics[scale=0.65]{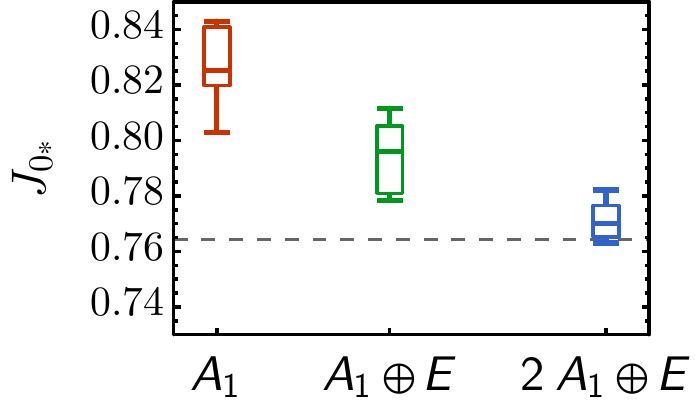}
\caption{Estimated critical point $J_{0*}$ under different choices of the on-site point group representation. The dashed line indicates the exact value $J_{0c}=\frac{1}{2}\mathrm{arccosh}(1+\sqrt{2})$.}
\label{fig: Jc}
\end{center}
\end{figure}

One can see that the RG monotone peaks at the critical point, and the peak becomes sharper with longer RG flow $\ell$. With a sufficiently large $\ell$, we can estimate the critical point $J_{0*}$ by finding the local maximum of the RG monotone
\eq{J_{0*}=\mathop{\mathrm{argmax}}_{J_0}C(J(J_0,\ell)),}
The result $J_{0*}$ will be insensitive to the RG scale $\ell$ as long as it is large enough that the RG flow has converged. We trained several different MLRG models for each on-site point group representation choice and estimated the Ising critical point $J_{0*}$ using the abovementioned method. Our result is shown in \figref{fig: Jc}. We can see a clear trend that the estimated critical point converges to its exact value $J_{0c}$ with larger point group representations (hence stronger RBM models).

\subsection{Boltzmann Weight Tensor}\label{sec: GSD}

\begin{figure}[htbp]
\begin{center}
\includegraphics[scale=0.65]{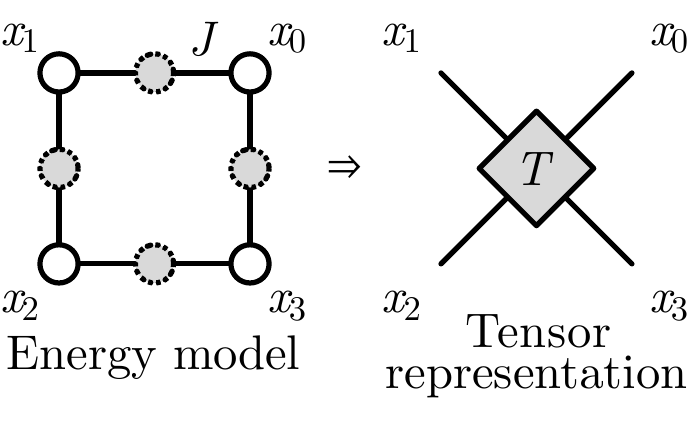}
\caption{Tensor representation of the RBM energy model, such that each tensor element encodes the Boltzmann weight of a configuration of visible spins.}
\label{fig: T}
\end{center}
\end{figure}

To further analyze the physical properties at different RG fixed points, we define the Boltzmann weight tensor $T(J)$ from the RBM energy model given the coupling parameter $J$. The tensor elements are specified as
\eq{\label{eq: def T} T(J)_{x_0,x_1,x_2,x_3}:=\sum_{z}\e^{-E_J(x,z)},}
where $x=(x_0,x_1,x_2,x_3)$ denotes the visible spins jointly and $z$ denotes the hidden spins jointly. $T(J)$ is a rank-4 tensor with each tensor element encoding the Boltzmann weight of a particular configuration of visible spins $x$, as illustrated in \figref{fig: T}.  We can use either the teacher RBM as \eqnref{eq: E tch} or the student RBM as \eqnref{eq: E std} for the energy model $E_J(x,z)$. They should not have much difference as long as the model parameter $J$ has converged to an RG fixed point where the teacher and the student should behave the same in the ideal limit. Nevertheless, in the following analysis, we will always adopt the teacher model as we found it a little more accurate than the student model.

The Boltzmann weight tensor $T(J)$ enables us to define various physical properties of the statistical mechanical model conveniently. For example, the ground state degeneracy  $Z(J)$ (regularized partition function) \cite{Gu2009T0903.1069,Yang2015L1512.04938}  can be defined as the ratio of the following tensor contractions (repeated indices are summed automatically)
\eq{Z(J)=\frac{(T(J)_{abab})^2}{T(J)_{acbc}T(J)_{bdad}}=\frac{\left(\raisebox{-15pt}{\includegraphics[scale=0.65]{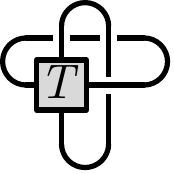}}\right)^2}{\includegraphics[scale=0.65]{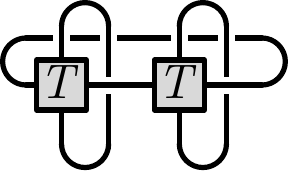}}.}
For Ising models, the ground state degeneracy characterizes the order of the broken symmetry group: $Z=|\dsZ_1|=1$ at the paramagnetic (disordered) fixed point where the internal $\dsZ_2$ symmetry is preserved, and $Z=|\dsZ_2|=2$ at the ferromagnetic (ordered) fixed point where the internal $\dsZ_2$ symmetry is broken.

\begin{figure}[htbp]
\begin{center}
\includegraphics[scale=0.65]{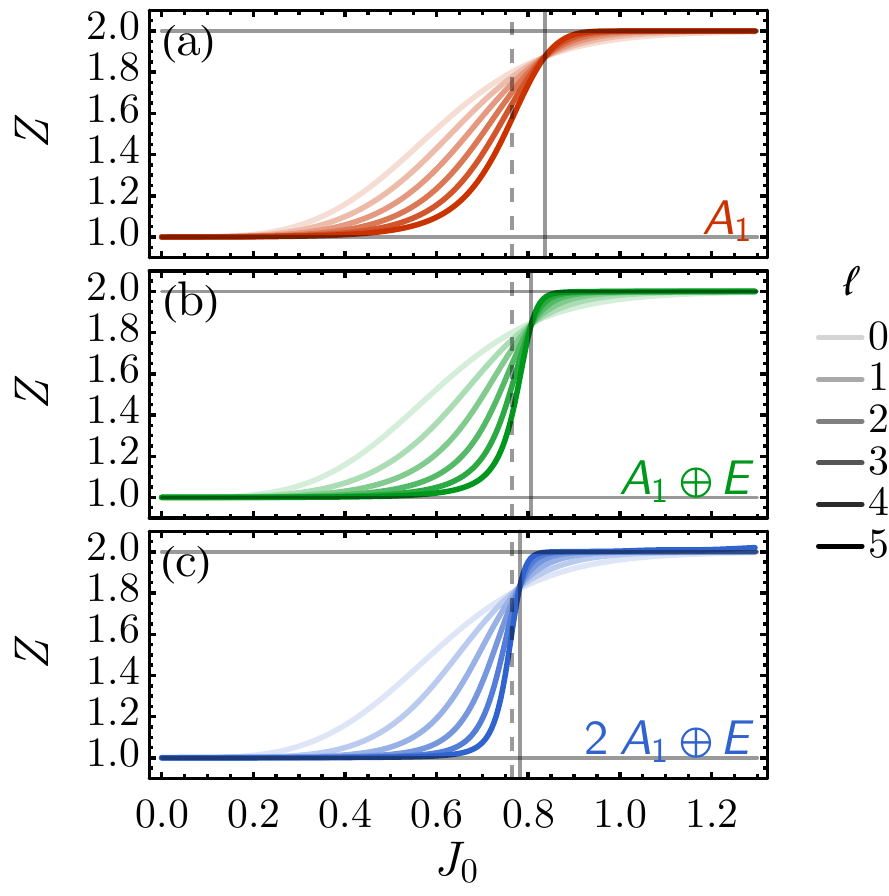}
\caption{Flow of the ground state degeneracy $Z$ starting from given $J_0$ parameters, under the representation choice of (a) $\mathsf{A}_1$, (b) $\mathsf{A}_1\oplus \mathsf{E}$, and (c) $2\mathsf{A}_1\oplus\mathsf{E}$. $\ell$ labels the steps of RG flow. The vertical solid line marks the estimated critical point $J_{0*}$, while the exact critical point $J_{0c}$ is displayed as the vertical dashed line.}
\label{fig: GSDflow}
\end{center}
\end{figure}

Following the approach described in \secref{sec: Jc}, we start with the initialization of the coupling matrix $J$ by a single parameter $J_0$ as in \eqnref{eq: J init} and follow the RG flow to obtain $J(J_0,\ell)$. At different RG scale $\ell$, we plot the ground state degeneracy $Z(J(J_0,\ell))$ as a function of the initial parameter $J_0$, the result is shown in \figref{fig: GSDflow}. As $J_0$ goes across the critial value $J_{0*}$, the ground state degeneracy transitions from $Z=1$ (paramagnet) to $Z=2$ (ferromagnet). The transition becomes sharper with increasing $\ell$ along the RG flow. The intersection of the ground state degeneracy curve at different RG scales can be used to infer the Ising critical point $J_{0*}$. As we enlarge the point group representation, $J_{0*}$ progressively approaches the exact Ising critical point at $J_{0c}$ as shown in \figref{fig: Jc}. 

The MLRG can automatically label different symmetry-breaking phases by their ground state degeneracy using this analysis. This can be viewed as an approach for unsupervised phase classification. The ground state degeneracy curves $Z(J(J_0,\ell))$ also provide a method to estimate the critical point $J_{0*}$ by finding their intersections.

\subsection{Locating the Monotone Saddle Point}\label{sec: Newton}

Our goal is to estimate the universal properties at the critical point. This requires us to accurately locate the RG fixed point parameter $J_*$ within the high-dimensional parameter space, as exemplified in \figref{fig: RG flow}.

However, if we simply follow the RG flow to approach $J_*$, it is almost inevitable that we will miss it. This is due to the tendency of the RG flow to deviate from the unstable fixed points. Fortunately, the MLRG algorithm has been trained to learn the RG monotone function $C(J)$. This feature enables us to find the RG fixed point by locating the saddle point of the RG monotone via Newton's method. By doing so, we can effectively circumvent the complication of fine-tuning the parameters in the parameter space.

\begin{figure}[htbp]
\begin{center}
\includegraphics[scale=0.65]{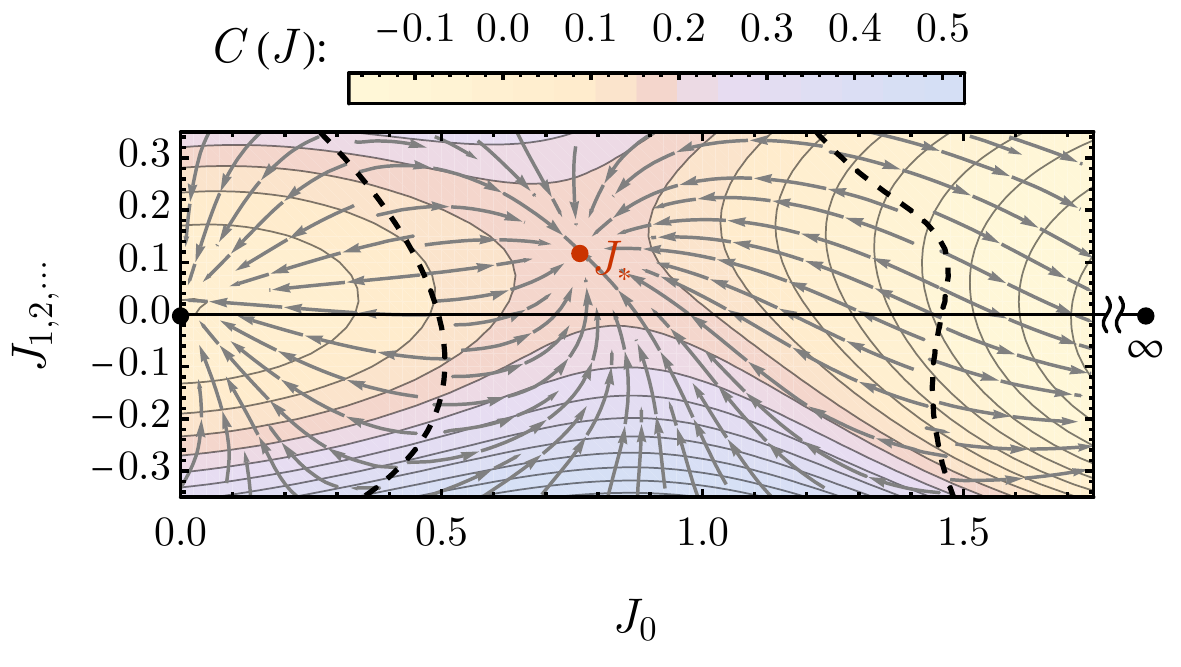}
\caption{Flow diagram of Newton's method iteration \eqnref{eq: Newton} to local RG fixed points for models with $\mathsf{A}_1\oplus \mathsf{E}$ representation. Both stable and unstable RG fixed points are attractive under Newton's flow. The dashed curves mark the borders between different attractive basins.}
\label{fig: newton}
\end{center}
\end{figure}

Given the RG monotone $C(J)$ and starting from an initial guess of $J$ near an RG fixed point, Newton's method recursively improves the approximation by the following update
\eq{\label{eq: Newton}
J\to J- (\nabla_J^2 C)^{-1}\nabla_J C,}
where $\nabla_J C$ denotes the gradient vector and $\nabla_J^2 C$ denotes the Hessian matrix of $C(J)$ at $J$. \figref{fig: newton} demonstrates the flow of $J$ along the vector field of $-(\nabla_J^2 C)^{-1}\nabla_J C$. Differing from the RG flow in \figref{fig: RG flow}, Newton's method converges to the RG fixed point in its neighborhood from all directions, regardless of whether the RG fixed point is stable or unstable. This enables us to pinpoint the RG fixed point from an initial guess near it. 

To locate the Ising critical fixed point $J_*$ in the parameter space, the initial guess can be constructed by setting the $J_0$ component of the matrix $J$ to its critical value $J_{0*}$, which can be estimated by methods described in \secref{sec: monotone} or \secref{sec: GSD}. The convergence can be monitored by evaluating $C(J)$. We stop the iteration when $C(J)$ does not change significantly between successive steps. The iteration is expected to converge to a saddle point $J_*$ of the RG monotone where $\nabla_J C=0$. 




\subsection{Calculating Scaling Dimensions}\label{sec: Delta}

Once we locate the RG fixed point $J_*$ by the iteration in \eqnref{eq: Newton}, we can substitute it into \eqnref{eq: def T} to construct the fixed-point Boltzmann weight tensor $T(J_*)$. The tensor can then be used to compute the transfer matrix $M$, with the following matrix elements
\eq{M_{ab}:=T(J_*)_{acbc}=\raisebox{-9pt}{\includegraphics[scale=0.65]{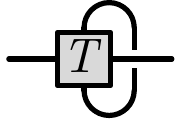}}.}
The eigenvalues of the transfer matrix $M$ are related to the scaling dimensions $\Delta_k$ ($k=0,1,\cdots$) of operators in the CFT  \cite{Gu2009T0903.1069,Yang2015L1512.04938,2023PhRvB.107t5123H,2023PhRvB.108b4413U,2021PhRvR...3b3048L},
\eq{\Tr M^n \propto \sum_{k}\e^{-2\pi n(\Delta_k-c/12)},}
with $c$ being the central charge. The proportionality factor can be fixed by requiring the identity operator (i.e., the vacuum state) to have zero scaling dimension $\Delta_0=0$. This allows us to extract the lowest few scaling dimensions associated with the most relevant operators.

\begin{figure}[htbp]
\begin{center}
\includegraphics[scale=0.65]{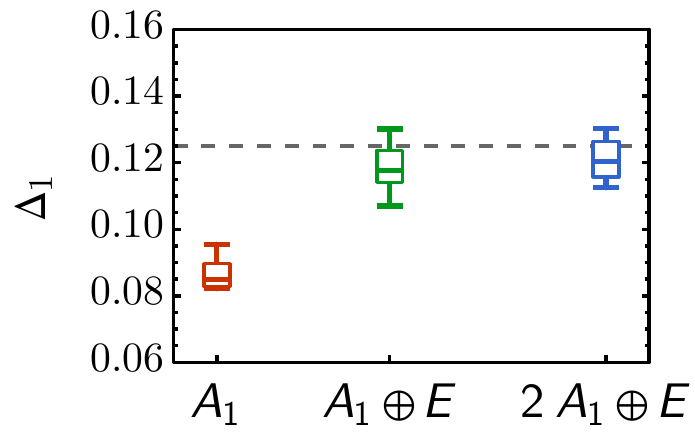}
\caption{Estimated scaling dimension $\Delta_1$ under different on-site point group representation choices. The dashed line indicates the exact value $\Delta_1=1/8$.}
\label{fig: scaling}
\end{center}
\end{figure}

\figref{fig: scaling} shows the estimation of the lowest scaling dimension $\Delta_1$ by the Ising MLRG with different point group representation choices. This should be the scaling dimension of the $\dsZ_2$-odd Ising order parameter, whose exact value is known to be $\Delta_1=1/8$. We can see that the MLRG can approach the exact value with larger point group representations.

\section{Summary and Discussion}\label{sec: summary}

In this study, we proposed the MLRG algorithm and used it to analyze many-body lattice models in statistical physics. Our method incorporates several modern machine learning techniques, including using equivariant neural networks \cite{Cohen2016G1602.07576,Kondor2018O1802.03690,Weiler201831807.02547,Cohen2018A1811.02017,Finzi2021A2104.09459,Lim2022W2205.07362} for RBMs and neural ordinary differential equations \cite{Chen2018N1806.07366,Grathwohl2018F1810.01367} for modeling the RG flow. The source code and raw data are available in the MLRG GitHub repository \cite{github}.

The MLRG algorithm demonstrates the power of machine learning to automate and enhance the study of statistical physics systems. We used the representation learning capability provided by generative modeling to learn the optimal RG transformation from the data directly, without requiring direct human intervention or prior knowledge about the system (apart from symmetry and dimensionality).

The design of the MLRG algorithm exemplifies the paradigm of \emph{introspective learning}  \cite{Wang2019E1901.11103}, an effective approach in machine learning for scientific discovery. It refers to the ability of the algorithm to self-analyze and self-adapt during the learning process, drawing on multiple levels of learning and analysis to generate a more comprehensive and insightful understanding of the problem at hand. MLRG's design incorporates this introspective nature through its multi-level, multi-model architecture. It uses two ``lower-level'' machines, the teacher and student models, to mimic the renormalization group process of extracting relevant features by representation learning. It also incorporates a ``higher-level'' machine, the moderator model, which learns the RG flow and uses its knowledge to guide the sampling of new model parameters that are most worth training. This multi-level, multi-model design allows for a clear separation between \emph{task performance}, carried out by the teacher and student models, and \emph{knowledge extraction}, carried out by the moderator model. This separation endows the high-level model with error-correcting capabilities, enabling it to resist the influence of noise in RBM training on learning the RG flow. This ability to compress knowledge and correct errors is crucial for AI to make scientific discoveries \cite{Raghu2020A2003.11755}.

The MLRG approach is related to (and distinguished from) existing works in the following ways:
\begin{itemize}
\item Monte Carlo Renormalization Group (MCRG): MCRG \cite{Ma1976R,Swendsen1984Ma,Pawley1984M,Gupta1984M,Swendsen1984Mb,Swendsen1984Mc,Baillie1992M,Blote1996Mcond-mat/9602020,Ron2002Ia,Ron2002Ib,Ron2017S1703.02430,Wu2017V1707.08683,Wu2019D1903.08231,Wu2019T1903.12137} uses Monte Carlo sampling to facilitate non-perturbative RG techniques. This process begins with sampling configurations from a fine-grained model given the model parameters. It then employs local RG transformations to generate coarse-grained configurations and estimates the new model parameters accordingly. This approach aligns with our method of training the student RBM from the teacher RBM. In particular, recent developments \cite{Chung2019O1912.09005,Chung2021N2010.05703,Ron2021M2011.05567} have introduced generative models to learn the optimal RG transformation, a goal that aligns with ours.

However, our method leverages specific lattice structures, such as the Lieb lattice, to restrict the RG operation within a small spin cluster. In the lack of this design, MCRG has to perform the sampling on a much larger lattice and model many long-range multi-spin couplings, which increases computational complexity. Furthermore, MCRG does not learn the RG monotone and can not automatically locate RG fixed points, such that it requires fine-tuning model parameters when studying critical point properties, adding to the computational burden.

\item Deep Learning and RG: Several studies \cite{Beny2013D1301.3124,Mehta2014A1410.3831,Beny2015T1402.4949,Lin2017W1608.08225,Shiba-Funai2018T1810.08179} have drawn parallels between deep neural networks and RG. Notably, they recognize that generation is the inverse process of renormalization \cite{Hu2020M1903.00804,Sheshmani2023C2203.07975}. Therefore, generative models can be used to implement data-driven RG. These discussions focus on optimizing RG transformations using deep learning, which often relies on Monte Carlo simulated spin configurations for hierarchical feature extraction. Despite their effectiveness in extracting features of stable phases, they lack controlled accuracy in predicting universal properties of phase transitions, as they did not learn the RG equation or the RG monotone.

\item RG Flow-Based Generative Modeling: Techniques such as Neural-RG \cite{Li2018N1802.02840,Hu2020M1903.00804} and RG-Flow \cite{Hu2022R2010.00029,Sheshmani2023C2203.07975} embed RG transformations in multi-level flow-based generative models \cite{Kingma2018G1807.03039,Kobyzev2019N1908.09257,Papamakarios2019N1912.02762}, applying deep learning methods to learn optimal RG transformations from model Hamiltonians by minimizing free energy. These methods are based on the \emph{invertible RG} framework, which designs the local RG transformation as a bijective (invertible) deterministic map from spin configurations to relevant and irrelevant features. 

However, the requirement for bijectivity limits the possibilities for RG transformation. Moreover, flow-based models struggle to model discrete variable probability distributions, limiting their application in various statistical mechanics problems. They also lack an asymptotic exact limit, meaning they can typically only serve as configuration update proposers to accelerate Monte Carlo calculations rather than replacing Monte Carlo as an unbiased simulation method. 

In contrast, MLRG applies to both discrete and continuous variables, offers more flexible RG transformations with stochastic coarse-graining maps, allows the extension of on-site degrees of freedom, and is exact when the on-site degrees of freedom tend to infinity. These characteristics make MLRG a valuable method for studying statistical physics.

\item Information Theoretical Approach to RG: Some research has explored the information theoretical criterion for optimal RG, indicating that RG transformation should maximize the mutual information of relevant features with the environment \cite{Koch-Janusz2018M1704.06279,Lenggenhager2018O1809.09632,Gordon2021R2012.01447,Gokmen2021S2101.11633,Gokmen2021S2103.16887,Gokmen2023C2301.11934} or minimize the mutual information among irrelevant features \cite{Hu2020M1903.00804}. While MLRG does not conflict with these principles, it does not explicitly use them as optimization criteria. Instead, it uses the match of marginal distributions of teacher and student models on their common spins to define optimal RG transformations, a method that is more direct and easier to optimize.
\end{itemize}

As a numerical algorithm to solve statistical mechanical problems, the MLRG showcases several advantages:
\begin{itemize}
\item Efficiency in Small Cluster Sampling: Compared to traditional Monte Carlo (MC) simulations, the MLRG algorithm operates on a smaller, lighter-weight scale. It only requires sampling spin configurations within a small cluster of spins (within a unit cell). Moreover, the Gibbs sampling in different spin clusters can be effectively parallelized on modern computing devices. This compact operation allows the algorithm to perform more efficiently than larger-scale simulations.

\item Exploration of the Full Parameter Space: The MLRG algorithm is designed to efficiently traverse the entire parameter space in a single pass of training. This differs significantly from MC simulations and Tensor Network Renormalization Group (TNRG) \cite{Levin2007Tcond-mat/0611687,Gu2008T0807.2010,Gu2009T0903.1069,Evenbly2015A1509.07484,Evenbly2015T1412.0732,Yang2015L1512.04938}, which must solve the Hamiltonian multiple times at different parameters. The MLRG algorithm's ability to explore the full parameter space reduces computational costs and time spent scanning the phase diagram, which can be beneficial when the parameter space dimension is large.

\item Discovery and Analysis of Critical Points: The MLRG makes use of the knowledge about the RG monotone to identify critical points (i.e., unstable RG fixed points). It uses machine-learned RG flow to calculate fixed-point properties, eliminating the need for finite-size scaling. This automated discovery and analysis process enhances the algorithm's capability to analyze critical properties in complex physical systems, especially for multi-critical points with multiple unstable directions.

\item Controlled Convergence of the Algorithm: Similar to tensor-network methods, the MLRG algorithm's estimation of physical properties converges as the dimension of the latent space increases. This behavior ensures that the algorithm's predictions become increasingly accurate and reliable as more computational resources are investigated.

\item Interpretability: Compared to other machine-learning methods for RG, the MLRG approach provides better interpretability by labeling the hidden spins in the model with symmetry representation, leading to physically meaningful coupling parameters and RG rules. This feature is particularly valuable in physical sciences, where the goal extends beyond accurate prediction to gaining physical insights. 
\end{itemize}

The current training of the RBMs that model local energy in MLRG is a stochastic process. This brings about fluctuations and noise, which can result in instability in the RG monotone network. The stability of the RG monotone network directly impacts the reliability of critical property predictions. A potential approach to improve this stability is to replace the RBMs with tensor networks, like in \figref{fig: T}. This could allow for a  deterministic training approach based on tensor network optimization, thus reducing the noise and increasing the stability of the RG monotone network. With the stability and reliability improvements, the MLRG can be extended to handle more complicated spin models. These models involve larger internal symmetry groups, such as $S_n$ groups, which are significant to the study of categorical symmetry \cite{Ji2019C1912.13492,Chatterjee2022E2212.14432} and entanglement transitions \cite{Vasseur2019E1807.07082,Li2018Q1808.06134,Skinner2019M1808.05953,Gullans2020D1905.05195,Bao2020T1908.04305,Jian2020M1908.08051}.

\begin{acknowledgments}
We acknowledge the discussions with Rokas Veitas, John McGreevy, Roger Melko, Han Ma, and Lei Wang. WH and YZY are supported by a startup fund by UCSD and the National Science Foundation (NSF) Grant No. DMR-2238360. The research was first presented at Swarma Club's reading group on Causal Emergence in 2021 summer and benefited from the interaction with the audience during the event. We acknowledge the OpenAI GPT4 model for providing editing suggestions throughout the process of writing this paper.
\end{acknowledgments}

\bibliographystyle{apsrev4-2}
\bibliography{ref}

\end{document}